\documentclass[a4paper,aps,prd,10pt,preprintnumbers,showpacs,onecolumn,superscriptaddress,nofootinbib,amsmath,amssymb,preprint]{revtex4-1}
\usepackage{graphicx,cmap,amsmath}
\usepackage[utf8]{inputenc}
\usepackage[T1]{fontenc}

\def\re#1{Re(#1)}
\def\im#1{Im(#1)}

\def\Order#1{{\cal O}\left(#1\right)}
\def\imo{i}

\newcommand{\eq}[1]{\begin{align} #1 \end{align}}

\begin{document}

\title{Gravitational perturbations of the Hayward spacetime and testing the correspondence between quasinormal modes and grey-body factors}

\author{Zainab Malik}\email{zainabmalik8115@outlook.com}
\affiliation{Institute of Applied Sciences and Intelligent Systems, H-15, Pakistan}

\begin{abstract}
The paper studies axial gravitational perturbations of the Hayward black hole, a regular geometry that also arises as an effective solution in asymptotically safe gravity. By computing grey-body factors with the 6th-order WKB method and comparing them to predictions based on the  quasinormal modes, the correspondence between transmission coefficients and quasinormal spectra is verified. Quantum corrections, parametrized by $\gamma$, are shown to suppress both the grey-body factors and the absorption cross-section, while the correspondence remains accurate at the percent level for low multipoles and essentially exact for higher ones.
\end{abstract}

\maketitle

\section{Introduction}

Quasinormal modes and grey-body factors represent two fundamental spectral characteristics of black holes. The former govern the damped oscillations in response to perturbations, encoding the ringdown phase of gravitational-wave signals \cite{Kokkotas:1999bd,Berti:2009kk,Konoplya:2011qq,Bolokhov:2025uxz}, while the latter quantify the frequency-dependent transmission probabilities of Hawking radiation  \cite{Hawking:1975vcx}  through the effective potential barrier surrounding the horizon \cite{Page:1976df,Page:1976ki,Kanti:2004nr}. Although defined by different boundary conditions, these quantities are not entirely independent. Recently, an approximate correspondence between quasinormal frequencies and grey-body factors was established \cite{Konoplya:2024lir}, showing that in the eikonal regime the transmission coefficients can be expressed in terms of the fundamental quasinormal modes, and that higher-order WKB refinements allow this relation to be extended to moderate multipole numbers. This observation is particularly interesting because grey-body factors, unlike overtones of the quasinormal spectrum \cite{Konoplya:2022pbc,Shen:2025yiy}, are much more stable under small deformations of the effective potential and thus may serve as robust probes of the underlying geometry \cite{Rosato:2024arw,Rosato:2025byu,Wu:2024ldo,Konoplya:2025ixm}. This correspondence was later generalized to rotating black holes \cite{Konoplya:2024vuj} and to wormhole geometries \cite{Bolokhov:2024otn}. It has also been tested and applied in a variety of other settings, including effective quantum-corrected black holes \cite{Skvortsova:2024msa,Konoplya:2024lch,Tang:2025mkk}, the Bonanno–Reuter metric \cite{Bolokhov:2025lnt}, Einstein–Gauss–Bonnet–Proca theory \cite{Lutfuoglu:2025ldc}, black holes surrounded by dark matter halos \cite{Hamil:2025pte}, Gibbons–Maeda–Garfinkle–Horowitz–Strominger black holes \cite{Dubinsky:2024vbn}, massive fields in Schwarzschild–de Sitter spacetime \cite{Malik:2024cgb}, and higher-dimensional black hole geometries \cite{Han:2025cal}.

The purpose of the present work is to explore this correspondence in the case of the Hayward black hole, which provides a widely studied example of a regular black hole and, upon a redefinition of constants, coincides with an effective geometry arising in asymptotically safe gravity. While the original motivation for the Hayward metric was to describe an evaporating black hole with a de Sitter core instead of a singularity, its modern interpretations extend far beyond this context, making it an ideal testbed for probing fundamental relations between quasinormal spectra and grey-body factors. By analyzing the transmission probabilities and quasinormal modes of the gravitational field in this background, we aim to verify the robustness of the correspondence proposed in \cite{Konoplya:2024lir} and to assess its applicability in geometries beyond the classical Schwarzschild spacetime. It should be noted that the quasinormal modes for gravitational perturbations were recently computed in Ref.~\cite{Bolokhov:2025egl}.  
The present work complements those results by analyzing the scattering properties of gravitational perturbations and by testing the correspondence between quasinormal modes and grey-body factors.

The paper is organized as follows. In Sec.~II we review the Hayward metric, discuss its relation to asymptotically safe gravity, and derive the master equation governing axial gravitational perturbations. In Sec.~III we compute the grey-body factors using the WKB method, test the correspondence with quasinormal modes, and evaluate the absorption cross-section. Finally, Sec.~IV summarizes our results and outlines possible directions for future work.

\section{The Hayward Metric and its Axial Gravitational Perturbations}\label{sec:hayward}

A central question in black-hole physics concerns the resolution of curvature singularities predicted by classical general relativity. One of the earliest and most influential proposals for a regular black hole geometry was given by Hayward~\cite{Hayward:2005gi}, who introduced a static, spherically symmetric metric designed to describe an evaporating black hole while avoiding the central singularity. The Hayward solution belongs to the class of so-called ``regular black holes,'' in which the metric interpolates smoothly between a Schwarzschild-like behavior at large distances and a de Sitter-like core at short scales, thereby replacing the singularity with a regular region of finite curvature.

The line element of the Hayward spacetime is
\begin{equation}
ds^2 = - f(r)\, dt^2 + \frac{dr^2}{f(r)} + r^2 \left( d\theta^2 + \sin^2\theta\, d\phi^2 \right),
\end{equation}
where the metric function is given by
\begin{equation}\label{fr}
f(r) = 1 - \frac{2 r^2 M}{r^3 + 2 M l^2}.
\end{equation}
Here $M$ denotes the mass parameter of the black hole, and $l$ is a new constant setting the scale at which quantum effects become relevant. For $r \gg (M l^2)^{1/3}$ the function reduces to the Schwarzschild form $f(r) \approx 1 - 2M/r$, while for $r \to 0$ it approaches the de Sitter-like core $f(r) \approx 1 - r^2/l^2$. Some properties of this spacetime have been considered in \cite{Filho:2023voz}.

Originally, the parameter $l$ was introduced phenomenologically, motivated by the idea that quantum backreaction or effective matter sources could regularize the black hole geometry. Later investigations, however, revealed that the Hayward form is not merely an ad hoc construction: up to a redefinition of constants, it coincides with the static, spherically symmetric quantum-corrected black hole solutions obtained within the framework of asymptotically safe gravity \cite{Held:2019xde},
\begin{equation}
    f(r) = 1 - \frac{2r^2/M^2}{r^3/M^3 + \gamma},
\end{equation}
where the parameter $\gamma$ governs the strength of the quantum correction. The condition
\[
\gamma \lessapprox \frac{32}{27}
\]
is required to ensure the existence of an event horizon. In this construction, the quantum-corrected geometry is derived by associating the cut-off parameter with the Kretschmann scalar \cite{Held:2019xde}.  
Alternative prescriptions for fixing the cut-off scale lead to different classes of quantum-corrected black hole spacetimes \cite{Bonanno:2000ep,Platania:2019kyx}. While quasinormal modes of test fields have been analyzed for several of these geometries \cite{Konoplya:2023aph,Konoplya:2022hll,Al-Badawi:2023lke,DuttaRoy:2022ytr,Lin:2013ofa}, the case of gravitational perturbations remains less explored, with the notable exception of the study of the Bonanno–Reuter metric in \cite{Bolokhov:2025lnt} and the effective metric obtained from quantum collapse \cite{Bonanno:2023rzk,Shi:2025gst,Bonanno:2025dry}. Gravitational quasinormal modes of the Hayward metric have been recently calculated in \cite{Bolokhov:2025egl}, as mentioned earlier. In the present work, we take the next step by investigating gravitational perturbations of the Hayward black hole, with particular attention to its interpretation in the context of asymptotically safe gravity.

Thus, the Hayward metric occupies an interesting dual role. On one hand, it serves as a prototypical example of a regular black hole spacetime free from central singularities, while on the other it can be regarded as an effective geometry emerging from a specific quantum gravity approach. This dual interpretation makes the Hayward spacetime a useful testing ground for probing potential quantum-gravity corrections to black hole observables within a fully relativistic framework.

The analysis of gravitational perturbations in this setting is considerably more subtle, since the background metric is obtained from an effective Hamiltonian-constraint approach rather than as an exact solution of Einstein’s equations with explicit quantum corrections. As a result, performing a fully rigorous treatment of perturbations is a difficult task. Nevertheless, as shown by Ashtekar, Olmedo, and Singh \cite{Ashtekar:2018lag,Ashtekar:2018cay}, quantum effects can be consistently mimicked by introducing an anisotropic fluid energy–momentum tensor within the Einsteinian framework. This reformulation makes it possible to analyze the dynamics of perturbations. Building on the approach of Bouhmadi-López and collaborators \cite{Bouhmadi-Lopez:2020oia,Konoplya:2024lch}, one can study axial perturbations under the simplifying assumption that fluctuations along the anisotropy direction do not contribute in the axial sector.  

This strategy mirrors the analysis of \cite{Bronnikov:2012ch}, where black holes coupled to scalar fields were considered and scalar perturbations decoupled from axial gravitational modes (see also \cite{Chen:2019iuo}). Such analogies strengthen confidence in the applicability of this approach, even if some sector-specific features may remain inaccessible.

In the Regge–Wheeler gauge \cite{Regge:1957td}, the axial gravitational perturbations $h_{\mu\nu}$ are parametrized as
\begin{widetext}
\begin{eqnarray}
h^{axial}_{\mu \nu}= \left[
 \begin{array}{cccc}
 0 & 0 &0 & h_0(t,r)
\\ 0 & 0 &0 & h_1(t,r)
\\ 0 & 0 &0 & 0
\\ h_0(t,r) & h_1(t,r) &0 &0
\end{array}\right]
\left(\sin\theta\frac{\partial}{\partial\theta}\right)
P_{\ell}(\cos\theta)\,, \label{pert_axial}
\end{eqnarray}
\end{widetext}
where $h_0(t,r)$ and $h_1(t,r)$ are two unknown functions, and $P_{\ell}(x)$ denotes the Legendre polynomial.  

Following \cite{Konoplya:2024lch}, the resulting perturbation equations can be interpreted as Einstein’s equations coupled to an anisotropic fluid with
\begin{equation}
    T_{\mu\nu}=(\rho+p_t)u_{\mu}u_{\nu}+g_{\mu\nu}p_t+(p_r-p_t)s_{\mu}s_{\nu},
\end{equation}
where $\rho$ is the energy density, $p_r$ and $p_t$ denote the radial and tangential pressures, respectively, and the fluid velocity $u_{\mu}$ together with the radial unit space-like vector $s_{\mu}$ are defined by
\begin{equation}
    u_{\mu}=(\sqrt{f(r)},0,0,0),\quad s_{\mu}=(0,1/\sqrt{f(r)},0,0),
\end{equation}
satisfying
\eq{
u_{\mu}u^{\mu}=-1,\quad s_{\mu}s^{\mu}=1,\quad u_{\mu}s^{\mu}=0.
}
Since $\rho$, $p_r$, and $p_t$ transform as scalars under the rotation group on the two-sphere, their axial perturbations vanish. For the vectors $u_{\mu}$ and $s_{\mu}$, the only nontrivial perturbed components are
\eq{
\delta u_{\phi}=-i\omega U(r)e^{-i\omega t}\sin\theta\partial_{\theta}P_{\ell}(\cos\theta),
}
\eq{
\delta s_{\phi}=-S(r)e^{-i\omega t}\sin\theta\partial_{\theta}P_{\ell}(\cos\theta).
}
Imposing the condition that there are no perturbations along the anisotropy direction, we further assume $\delta s_{\mu}=0$.  

From the conservation equation $\nabla_{\mu}T^{\mu r}=0$, one immediately finds $\delta u_{\phi}=0$. Substituting these relations into Einstein’s equations yields
\begin{align}
        & h_1(r) \left(r^2 \omega^2 - (\ell - 1)(\ell + 2) f(r)\right) \nonumber \\
        &\quad - i r^2 \omega h_0'(r) + 2 i r \omega h_0(r) = 0,
\end{align}
together with
\eq{
f(r) \left(\frac{h_1(r) f'(r)}{f(r)}+2 h_1'(r)\right)+\frac{2 i
   \omega  h_0(r)}{g(r)}=0.
}
Here $\ell=2,3,4,\ldots$ is the multipole number.

After some algebra, and introducing the new variables,
\eq{h_1=\frac{r}{f(r)}\Psi,\quad dr^*=\frac{dr}{f(r)},}
the perturbation equation takes the Schrödinger-like form
\eq{
\frac{d^2}{dr_{*}{}^2}\Psi+\left(\omega^2-V(r)\right)\Psi=0,
}
with the effective axial potential
\eq{
V=f(r)\left(\frac{2f(r)}{r^2}-\frac{f'(r)}{r}+\frac{(\ell+2)(\ell-1)}{r^2}
\right).
}

Similar simplifications have been employed in a number of related contexts, where certain perturbations are assumed negligible and consequently omitted \cite{Berti:2003yr,Kokkotas:1993ef,Konoplya:2006ar}. While this procedure may omit potentially interesting corrections to the full gravitational spectrum, it nonetheless provides a good first approximation when deviations from the Schwarzschild geometry remain small. This viewpoint is consistent with the perturbative nature of quantum corrections, which are expected to affect the spacetime only modestly.  

\begin{figure}
\resizebox{\linewidth}{!}{\includegraphics{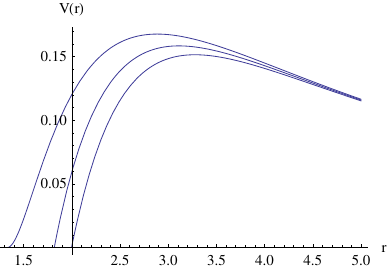}~~\includegraphics{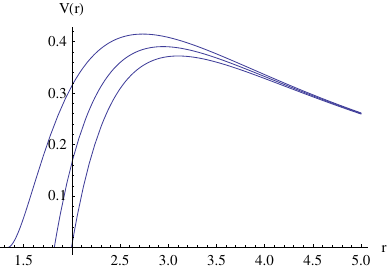}}
\caption{Effective potentials for $\ell=2$ (left) and $\ell=3$ (right) axial gravitational perturbations: $M=1$, $\gamma=0.01$ (bottom), $\gamma=0.6$ (middle), and $\gamma=32/27$ (top).}\label{fig:Pot}
\end{figure}

\section{Grey-Body Factors and Testing the Correspondence}

We begin by illustrating the effective potential for gravitational perturbations as a function of the radial coordinate for different values of the quantum parameter $\gamma$. Figure~\ref{fig:Pot} shows that increasing $\gamma$ leads to a smaller event horizon radius and a higher peak of the effective potential. At distances around $r \gtrsim 5M$, the metrics for different values of $\gamma$ practically coincide with the Schwarzschild case. Therefore, the influence of quantum corrections is confined primarily to the near-horizon region.

\begin{figure}
\resizebox{\linewidth}{!}{\includegraphics{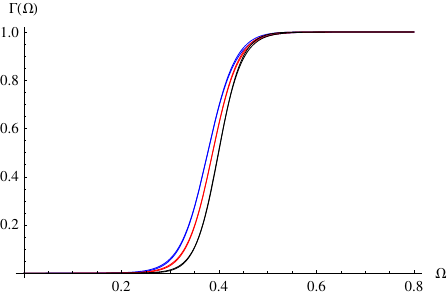}~~\includegraphics{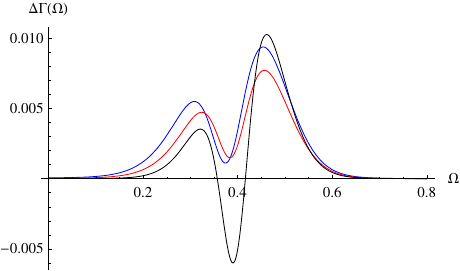}}
\caption{Grey-body factors of $\ell=2$ axial gravitational perturbations calculated by the 6th order WKB method and via the correspondence with QNMs: $M=1$, $\gamma=0.01$ (top, blue), $\gamma=0.6$ ( middle, red), and $\gamma=32/27$ (bottom, black).}\label{fig:L2}
\end{figure}
\begin{figure}
\resizebox{\linewidth}{!}{\includegraphics{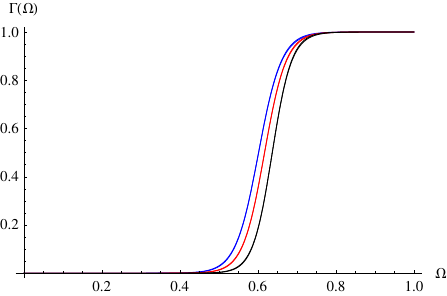}~~\includegraphics{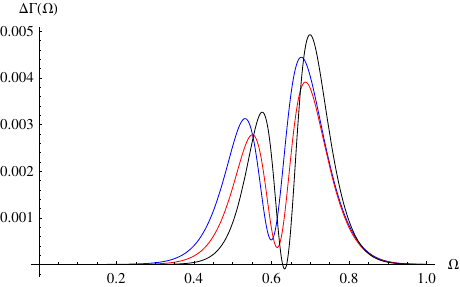}}
\caption{Grey-body factors of $\ell=3$ axial gravitational perturbations calculated by the 6th order WKB method and via the correspondence with QNMs: $M=1$, $\gamma=0.01$ (top, blue), $\gamma=0.6$ ( middle, red), and $\gamma=32/27$ (bottom, black).}\label{fig:L3}
\end{figure}
\begin{figure}
\resizebox{\linewidth}{!}{\includegraphics{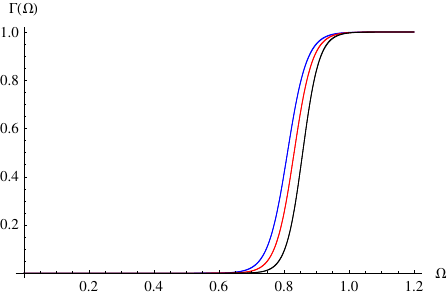}~~\includegraphics{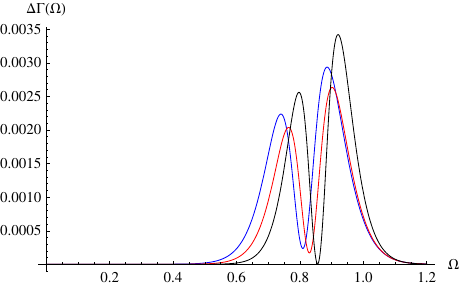}}
\caption{Grey-body factors of $\ell=4$ axial gravitational perturbations calculated by the 6th order WKB method and via the correspondence with QNMs: $M=1$, $\gamma=0.01$ (top, blue), $\gamma=0.6$ ( middle, red), and $\gamma=32/27$ (bottom, black).}\label{fig:L4}
\end{figure}
\begin{figure}
\resizebox{\linewidth}{!}{\includegraphics{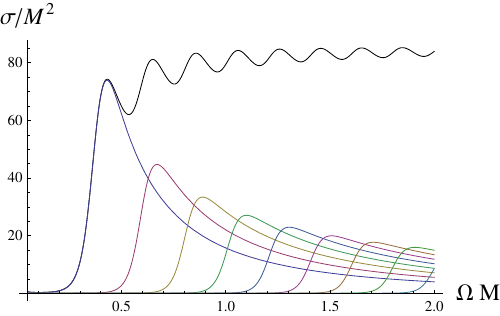}~~\includegraphics{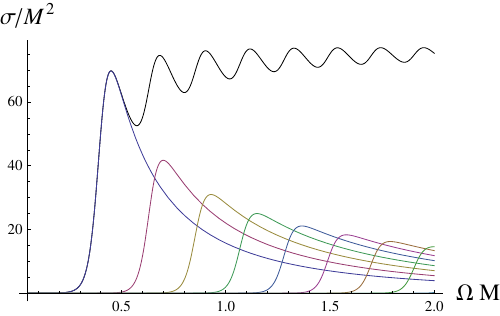}}
\caption{Absorption cross-section for gravitons for the first ten multipoles $\ell=2,3,...11$ together with the total cross-section (black curve) } at $\gamma =0.01$ (left) and $\gamma=32/27$ (right). The suppression of the cross-section for the near extreme black hole is about $5\%$.\label{fig:Sigma}
\end{figure}

Hawking radiation emitted by black holes is often described as thermal, yet this description is only approximate.  
In reality, the radiation spectrum detected at infinity is shaped by the effective potential barrier surrounding the black hole.  
Only part of the radiation created near the horizon manages to penetrate this barrier and escape to infinity, while the rest is reflected back.  
The deviation from an exact blackbody spectrum is encoded in the \emph{grey-body factors}, which depend upon the geometry of spacetime, the nature of the perturbing field, and the underlying gravitational theory~\cite{Page:1976df,Page:1976ki,Kanti:2004nr}.  

For a wave of frequency $\Omega$ and multipole index $\ell$, the grey-body factor $\Gamma_{\ell}(\Omega)$ quantifies the transmission probability through the potential barrier.  
To define it precisely, one considers asymptotic boundary conditions for the radial wave function:
\begin{eqnarray}
\Psi(r_{*}\to -\infty) &=& A_{\text{trans}} e^{-i\Omega r_{*}}, \\
\Psi(r_{*}\to +\infty) &=& A_{\text{out}}\,e^{+i\Omega r_{*}}
+ A_{\text{in}}\,e^{-i\Omega r_{*}}.
\end{eqnarray}
Thus, in the scattering problem, the purely ingoing condition at the horizon is imposed, while at infinity the solution is a superposition of incident and reflected waves.  
The grey-body factor then follows as
\begin{equation}
\Gamma_{\ell}(\Omega) = 
\left|\frac{A_{\text{trans}}}{A_{\text{in}}}\right|^{2}
= 1 - \left|\frac{A_{\text{out}}}{A_{\text{in}}}\right|^{2},
\end{equation}
with $A_{\text{in}}$, $A_{\text{out}}$, and $A_{\text{trans}}$ representing the amplitudes of the incoming, reflected, and transmitted waves, respectively.
Here, $\Omega$ is a real frequency and should not be confused with the quasinormal modes $\omega$.

The quasinormal modes boundary conditions are different and require purely ingoing wave at the event horizon and purely outgoing one at infinity:
\begin{eqnarray}
\Psi(r_*) &\propto& e^{-i \omega r_*}, \quad r_* \to -\infty  \quad (\text{event horizon}),\\
\Psi(r_*) &\propto& e^{+i \omega r_*}, \quad r_* \to +\infty \quad (\text{infinity}) .
\end{eqnarray}

Although direct numerical integration of the wave equation provides highly accurate results for $\Gamma_{\ell}(\Omega)$, in practice semi-analytical techniques are often more convenient.  Among these, the Wentzel–Kramers–Brillouin (WKB) method has become one of the most widely used tools for black hole scattering problems.  
Originally adapted for quasinormal mode calculations in~\cite{Schutz:1985km,Iyer:1986np,Konoplya:2003ii}, the WKB approach matches asymptotic solutions across the classically forbidden region near the barrier, yielding approximate expressions for both reflection and transmission coefficients.  

In its standard implementation, the WKB approximation is carried out around the maximum of the effective potential. The transmission probability is given by \cite{Schutz:1985km}
\begin{equation}
\Gamma_{\ell}(\Omega) = 
\frac{1}{1 + \exp\!\left(2\pi i K\right)},
\end{equation}
where the quantity $K$ depends on the barrier height $V_{0}$ and its curvature $V_{0}''$ at the maximum.  
Corrections at higher orders involve higher derivatives of the potential \cite{Matyjasek:2017psv,Konoplya:2019hlu}.  
In the present work we restrict ourselves to the 6th-order WKB expansion~\cite{Konoplya:2003ii}. This framework has been successfully employed in a broad range of contexts, from classical general relativity to higher-derivative theories (see~\cite{Miyachi:2025ptm, Matyjasek:2017psv, Matyjasek:2021xfg,Matyjasek:2021xfg,Pedrotti:2025idg,Konoplya:2019hlu,Lutfuoglu:2025ljm,Hamil:2025cms,Konoplya:2020cbv,MahdavianYekta:2019pol,Konoplya:2023moy,Konoplya:2021ube,Antonelli:2025yol,Zhang:2025xqt} for recent extensions and applications).  

A recent development has provided a new perspective on grey-body spectra.  
It was shown in~\cite{Konoplya:2024lir} that the grey-body factors are intimately related to the lowest quasinormal modes of the black hole.  
In the eikonal regime, one can approximate the transmission coefficient as
\[
\Gamma_{\ell}(\Omega) \approx \left[ 1 + \exp\!\left( \frac{2\pi\bigl(\Omega^{2}-\mathrm{Re}(\omega_{0})^{2}\bigr)}{4\,\mathrm{Re}(\omega_{0})\,\mathrm{Im}(\omega_{0})} \right) \right]^{-1},
\]
where $\omega_{0}$ denotes the fundamental quasinormal frequency.  
For intermediate multipoles, the precision improves if the first overtone $\omega_{1}$ is included through the correction
\[
-iK = -\frac{\Omega^{2}-\mathrm{Re}(\omega_{0})^{2}}{4\,\mathrm{Re}(\omega_{0})\,\mathrm{Im}(\omega_{0})} +  \frac{\mathrm{Re}(\omega_{0})-\mathrm{Re}(\omega_{1})}{16\,\mathrm{Im}(\omega_{0})}.
\]
Further refinements may involve additional corrections, which are explicitly written in \cite{Konoplya:2024lir}. An important feature of all these correcting terms is that they all depend only on the fundamental mode and first overtone and do not depend on higher overtones or multipole number $\ell$.  Here we used all the correcting terms presented in \cite{Konoplya:2024lir}, namely
\begin{eqnarray}\nonumber
&&\imo K=\frac{\Omega^2-\re{\omega_0}^2}{4\re{\omega_0}\im{\omega_0}}\Biggl(1+\frac{(\re{\omega_0}-\re{\omega_1})^2}{32\im{\omega_0}^2}
\\\nonumber&&\qquad\qquad-\frac{3\im{\omega_0}-\im{\omega_1}}{24\im{\omega_0}}\Biggr)
-\frac{\re{\omega_0}-\re{\omega_1}}{16\im{\omega_0}}
\\\nonumber&& -\frac{(\Omega^2-\re{\omega_0}^2)^2}{16\re{\omega_0}^3\im{\omega_0}}\left(1+\frac{\re{\omega_0}(\re{\omega_0}-\re{\omega_1})}{4\im{\omega_0}^2}\right)
\\\nonumber&& +\frac{(\Omega^2-\re{\omega_0}^2)^3}{32\re{\omega_0}^5\im{\omega_0}}\Biggl(1+\frac{\re{\omega_0}(\re{\omega_0}-\re{\omega_1})}{4\im{\omega_0}^2}
\\\nonumber&&\qquad +\re{\omega_0}^2\Biggl(\frac{(\re{\omega_0}-\re{\omega_1})^2}{16\im{\omega_0}^4}
\\&&\qquad\qquad -\frac{3\im{\omega_0}-\im{\omega_1}}{12\im{\omega_0}}\Biggr)\Biggr)+ \Order{\frac{1}{\ell^3}}.
\label{eq:gbsecondorder}
\end{eqnarray}

The quasinormal modes of test fields in the Hayward spacetime were studied in \cite{Konoplya:2022hll,Malik:2024tuf}, and for gravitational perturbations within the framework of Effective Field Theory in  \cite{Konoplya:2023ppx,Takahashi:2019oxz}, where the perturbation equations and the underlying assumptions differ completely from those of the Asymptotic Safety approach. In the Hayward spacetime as an effective metric in the Asymptotically Safe Gravity quasinormal modes were found in \cite{Bolokhov:2025egl} and here we confirm their results. To determine the quasinormal modes $\omega_0$ and $\omega_1$, we employed the sixth-order WKB method with Padé approximants $\tilde{m} = \tilde{n} = 3$, as recommended in \cite{Konoplya:2019hlu}, which provides the best accuracy in the vast majority of cases \cite{Bolokhov:2024ixe,Dubinsky:2025fwv,Konoplya:2019xmn,Lutfuoglu:2025hjy,Dubinsky:2024nzo,Konoplya:2020jgt,Malik:2024nhy,Malik:2023bxc,Bronnikov:2019sbx,Bolokhov:2023bwm,Skvortsova:2024atk,Skvortsova:2024wly,Skvortsova:2023zmj,Churilova:2021tgn}.

From figs.~\ref{fig:L2} - \ref{fig:L4} one can see that increasing the parameter $\gamma$ reduces the grey-body factors.  
This behavior is naturally explained by the effective potential (fig.~\ref{fig:Pot}), which grows higher as $\gamma$ increases, thereby suppressing the transmission probability and allowing a smaller fraction of particles to cross the barrier.  
Consequently, quantum corrections tend to decrease the Hawking evaporation rate at a fixed Hawking temperature.  
Nevertheless, the temperature itself remains the dominant factor determining the overall intensity of Hawking radiation, while the role of the grey-body factors is usually subleading.  
There are, however, examples where the grey-body suppression is nearly as important as the temperature in shaping the evaporation rate~\cite{Konoplya:2019ppy}.

We can also see that the relative error of the correspondence remains within about one to two percent for $\ell=2$, and below one percent for $\ell=3$. For $\ell=4$, the relative error is only a small fraction of one percent and becomes practically negligible for higher multipoles. 

It should be emphasized that the correspondence performs poorly when the effective potential possesses more than one peak, since the effects of secondary scattering are not taken into account by the WKB method on which the correspondence relies \cite{Konoplya:2025ixm}. This occurs also in the same situations where the correspondence between null geodesics and eikonal quasinormal modes breaks down, such as in certain higher-curvature theories \cite{Konoplya:2020bxa,Konoplya:2017wot,Konoplya:2025afm,Bolokhov:2023dxq} or in asymptotically de Sitter spacetimes \cite{Konoplya:2022gjp,Konoplya:2025mvj}. For the effective potential considered here, however, the WKB method applies reliably, which explains the high accuracy of the correspondence we have observed.
 
Using the data for grey-body factors for various multipole numbers, we can find
the absorption cross-section \cite{Futterman:1988ni},
\[
\sigma(\Omega)
 \;=\; \frac{\pi}{\Omega^{2}} \sum_{\ell=2}^{\infty} (2\ell+1)\,\Gamma_{\ell}(\Omega)
\]
The absorption cross-section $\sigma(\Omega)$ represents the effective area of the black hole that interacts with an incoming wave of frequency $\Omega$. Physically, it quantifies the probability that incident radiation is transmitted through the potential barrier and absorbed by the black hole, rather than being scattered back to infinity (see recent examples in \cite{OuldElHadj:2025hbl,C:2024cnk,Li:2024xyu,Heidari:2024bkm,Polo:2024xqm}).  

In fig. \ref{fig:Sigma} one can observe noticeable suppression of the absorption cross-section, when the quantum correction given by the parameter $\gamma$ is increased.

\section*{IV. CONCLUSIONS}

In this work we have analyzed axial gravitational perturbations of the Hayward black hole, which may be regarded both as a prototype of a regular black hole and as an effective geometry arising in asymptotically safe gravity. Using the sixth-order WKB approximation, we computed grey-body factors and compared them with the predictions obtained from the fundamental quasinormal modes via the recently proposed correspondence.  

Our results show that increasing the quantum correction parameter $\gamma$ raises the height of the effective potential barrier, leading to a suppression of the transmission probabilities and a corresponding reduction of the absorption cross-section. At the same time, the correspondence between grey-body factors and quasinormal frequencies has been confirmed with high accuracy: the relative error does not exceed a few percent for low multipoles and becomes practically negligible for higher ones.  

These findings indicate that the correspondence provides a robust link between scattering and oscillation properties of black holes even in the presence of quantum-inspired modifications of the geometry. Future work could extend this analysis to polar gravitational perturbations, rotating regular black holes, and more general classes of quantum-corrected spacetimes.

\textbf{Funding Declaration.} Funding information - not applicable.

\bibliography{bibliography}

\end{document}